\documentclass[a4paper]{jpconf}
\pdfoutput=1
\usepackage[centertags]{amsmath}
\usepackage{graphicx}
\usepackage{cite}
\usepackage{subfigure}
\usepackage{multirow}
\usepackage{bigdelim}
\begin{document}

\title{Precise Determination of the ${}^{235}\text{U}$ Reactor Antineutrino Cross Section per Fission}

\author{Carlo Giunti}

\address{INFN, Sezione di Torino, Via P. Giuria 1, I--10125 Torino, Italy}

\ead{giunti@to.infn.it}

\begin{abstract}
We consider the possibility that
the reactor antineutrino anomaly
is due to a miscalculation of one or more of the
$^{235}\text{U}$,
$^{238}\text{U}$,
$^{239}\text{Pu}$, and
$^{241}\text{Pu}$
reactor antineutrino fluxes.
From the fit of the data we obtain the precise determination
$\sigma_{f,235} =
( 6.33 \pm 0.08 )
\times 10^{-43} \, \text{cm}^2 / \text{fission}$
of the $^{235}\text{U}$ cross section per fission,
which is more precise than the
calculated value and differs from it
by $2.2\sigma$.
The cross sections per fission of the other fluxes
have large uncertainties and in practice their values
are undetermined by the fit.
We conclude that
it is very likely that at least the calculation of the $^{235}\text{U}$ flux
must be revised.
\end{abstract}

The reactor antineutrino anomaly
\cite{Mention:2011rk}
is due to the 2011 recalculation
\cite{Mueller:2011nm,Huber:2011wv}
of the reactor antineutrino flux,
which is about 3\% higher than the previous estimate
\cite{Schreckenbach:1985ep,Hahn:1989zr}
and
implies a deficit of the rate of $\bar\nu_{e}$ observed in several
reactor neutrino experiments.

%Electron antineutrinos are produced in nuclear reactors
%by the $\beta$ decays of the fission products of
%$^{235}\text{U}$,
%$^{238}\text{U}$,
%$^{239}\text{Pu}$, and
%$^{241}\text{Pu}$.
%The calculation of the antineutrino flux is based on the inversion
%of the spectra of the electrons
%emitted by the $\beta$ decays of the products of the thermal fission of
%$^{235}\text{U}$,
%$^{239}\text{Pu}$, and
%$^{241}\text{Pu}$
%which have been measured at ILL in the 80's
%\cite{Schreckenbach:1985ep,Hahn:1989zr,Haag:2014kia}.
%Since the fission of $^{238}\text{U}$
%is induced by fast neutrons,
%the measurement of its $\beta$ spectrum
%is more difficult and it was performed only recently
%at the scientific neutron source FRM II in Garching
%\cite{Haag:2013raa}.
%The $^{238}\text{U}$ antineutrino spectrum obtained from the conversion
%is about 10\% below that
%calculated in Ref.~\cite{Mueller:2011nm}
%for antineutrino energies between about 4.5 and 6.5 MeV.
%However, using the $^{238}\text{U}$ antineutrino spectrum
%of Ref.~\cite{Haag:2013raa}
%cannot solve the reactor antineutrino anomaly
%because:
%a)
%the contribution of
%$^{238}\text{U}$ in the neutrino experiments using
%highly enriched $^{235}\text{U}$ research reactors
%is negligible;
%b)
%for neutrino experiments using commercial reactors
%the contribution of
%$^{238}\text{U}$ to the total antineutrino flux is about 8\%
%and the change of the total integrated flux is only about 0.2\%
%\cite{An:2016srz}.

It is possible that
the reactor antineutrino anomaly is due to
the oscillations of the reactor $\bar\nu_{e}$'s
into sterile neutrinos with a mass at the eV scale
(see the review in Ref.~\cite{Gariazzo:2015rra}).
However,
it is also possible that the reactor antineutrino anomaly
is due to a flaw in the calculation
of one or more of the
$^{235}\text{U}$,
$^{238}\text{U}$,
$^{239}\text{Pu}$, and
$^{241}\text{Pu}$ fluxes
that compose the reactor antineutrino flux.
In this paper we consider this second possibility
and we investigate which of the four fluxes could be the cause of the
reactor antineutrino anomaly
\cite{Giunti:2016elf}.

The prime suspect as a cause for the reactor antineutrino anomaly
is the $^{235}\text{U}$ antineutrino flux,
because some of the experiments which observed a deficit of
electron antineutrinos
used research reactors, which produce an almost pure
$^{235}\text{U}$
antineutrino flux.
However,
since other experiments used commercial reactors
with significant contributions of the
$^{238}\text{U}$,
$^{239}\text{Pu}$, and
$^{241}\text{Pu}$
electron antineutrino fluxes,
a detailed calculation is necessary
in order to reach a definite and quantitative conclusion.

The theoretical prediction for the event rate of an experiment
labeled with the index $a$
is usually expressed by the cross section per fission
$
\sigma_{f,a}
=
\sum_{k} f^{a}_{k} \sigma_{f,k}
$,
with
$k = 235, 238, 239, 241$.
Here
$f^{a}_{k}$ is the antineutrino flux fraction from the fission of the
isotope with atomic mass $k$
and
$\sigma_{f,k}$
is the corresponding cross section per fission,
which is given by the integrated product of the antineutrino flux and the detection cross section.

The cross sections per fission of the four fissile isotopes calculated
by the Saclay group in Ref.~\cite{Mention:2011rk}
are
listed in Table~\ref{tab:csf}.
These values must be increased by
1.2\%,
1.4\%, and
1.0\%
for
${}^{235}\text{U}$,
${}^{239}\text{Pu}$, and
${}^{241}\text{Pu}$,
respectively,
according to the improved inversion of the ILL electron spectra of Huber
\cite{Huber:2011wv}.
The resulting values listed in Table~\ref{tab:csf}
coincide with those given in Table XX of Ref.~\cite{Abazajian:2012ys}.

\begin{table}[t!]
\begin{center}
\begin{tabular}{cccc}
&
Saclay (S)
&
Saclay+Huber (SH)
&
uncertainty
\\
\hline
$\sigma_{f,235}$ &  6.61 &  6.69 & 2.11\% \\
$\sigma_{f,238}$ & 10.10 & 10.10 & 8.15\% \\
$\sigma_{f,239}$ &  4.34 &  4.40 & 2.45\% \\
$\sigma_{f,241}$ &  5.97 &  6.03 & 2.15\% \\
\hline
\end{tabular}
\end{center}
\caption{ \label{tab:csf}
Cross sections per fission of the four fissile isotopes calculated
by the Saclay (S) group in Ref.~\cite{Mention:2011rk}
and those obtained from the Huber (SH) correction in Ref.~\cite{Huber:2011wv}.
The units are $10^{-43} \, \text{cm}^2 / \text{fission}$.
The uncertainties are those estimated by the
Saclay group in Ref.~\cite{Mention:2011rk}.
}
\end{table}

The experiments which measured the absolute antineutrino flux
are listed in Table~\ref{tab:rat}.
For each experiment labeled with the index $a$,
we listed the corresponding four fission fractions $f^{a}_{k}$,
the ratio of measured and predicted rates $R_{a,\text{SH}}^{\text{exp}}$,
the corresponding relative experimental uncertainty
$\sigma_{a}^{\text{exp}}$,
and
the relative uncertainty
$\sigma_{a}^{\text{cor}}$
which is correlated in each group of experiments
indicated by the braces.

For the short-baseline experiments
(Bugey-4 \cite{Declais:1994ma},
Rovno91 \cite{Kuvshinnikov:1990ry},
Bugey-3 \cite{Declais:1995su},
Gosgen \cite{Zacek:1986cu},
ILL \cite{Kwon:1981ua,Hoummada:1995zz},
Krasnoyarsk87 \cite{Vidyakin:1987ue},
Krasnoyarsk94 \cite{Vidyakin:1990iz,Vidyakin:1994ut},
Rovno88 \cite{Afonin:1988gx},
SRP \cite{Greenwood:1996pb}),
we calculated the
Saclay+Huber
ratios $R_{a,\text{SH}}^{\text{exp}}$
by rescaling the corresponding Saclay value
$R_{a,\text{S}}^{\text{exp}}$
in Ref.~\cite{Mention:2011rk}:
\begin{equation}
R_{a,\text{SH}}^{\text{exp}}
=
R_{a,\text{S}}^{\text{exp}}
\dfrac{\sum_{k} f^{a}_{k} \sigma_{f,k}^{\text{S}}}{\sum_{k} f^{a}_{k} \sigma_{f,k}^{\text{SH}}}
\qquad
(a=1,\ldots,17,19,20)
.
\label{rescaling}
\end{equation}
We considered the Krasnoyarsk99-34 experiment \cite{Kozlov:1999ct}
that was not considered in Refs.~\cite{Mention:2011rk,Zhang:2013ela},
by rescaling the value of the corresponding experimental cross section per fission
in comparison with the Krasnoyarsk94-57 result.
For the long-baseline experiments
Chooz \cite{Apollonio:2002gd}
and
Palo Verde \cite{Boehm:2001ik},
we applied the rescaling
with the ratios $R_{a,\text{S}}^{\text{exp}}$
given in Ref.~\cite{Zhang:2013ela},
divided by the corresponding
survival probability $P_{\text{sur}}$
caused by $\vartheta_{13}$.
For
Nucifer
\cite{Boireau:2015dda}
Daya Bay
\cite{An:2016srz},
RENO
\cite{RENO-Nu2016,SBKim-private-16},
and
Double Chooz
\cite{DoubleChooz-private-16}
we use the ratios provided by the respective experimental collaborations.

\begin{table}[t!]
\begin{center}
\begin{tabular}{cccccccccc}
$a$
&
Experiment
&
$f^{a}_{235}$
&
$f^{a}_{238}$
&
$f^{a}_{239}$
&
$f^{a}_{241}$
&
$R_{a,\text{SH}}^{\text{exp}}$
&
$\sigma_{a}^{\text{exp}}$ [\%]
&
$\sigma_{a}^{\text{cor}}$ [\%]
&
$L_{a}$ [m]
\\
\hline
1 	&Bugey-4 		 &0.538	&0.078	&0.328	&0.056	&0.932	&1.4	&\rdelim\}{2}{20pt}[1.4]	&$15$\\
2 	&Rovno91 		 &0.606	&0.074	&0.277	&0.043	&0.930	&2.8	&                       	&$18$\\
\hline
3 	&Rovno88-1I 		 &0.607	&0.074	&0.277	&0.042	&0.907	&6.4	&\rdelim\}{2}{20pt}[3.1] \rdelim\}{5}{20pt}[2.2]	&$18$\\
4 	&Rovno88-2I 		 &0.603	&0.076	&0.276	&0.045	&0.938	&6.4	&                                               	&$18$\\
5 	&Rovno88-1S 		 &0.606	&0.074	&0.277	&0.043	&0.962	&7.3	&\rdelim\}{3}{45pt}[3.1]                        	&$18$\\
6 	&Rovno88-2S 		 &0.557	&0.076	&0.313	&0.054	&0.949	&7.3	&                                               	&$25$\\
7 	&Rovno88-2S 		 &0.606	&0.074	&0.274	&0.046	&0.928	&6.8	&                                               	&$18$\\
\hline
8 	&Bugey-3-15 		 &0.538	&0.078	&0.328	&0.056	&0.936	&4.2	&\rdelim\}{3}{20pt}[4.0]                        	&$15$\\
9 	&Bugey-3-40 		 &0.538	&0.078	&0.328	&0.056	&0.942	&4.3	&                                               	&$40$\\
10 	&Bugey-3-95 		 &0.538	&0.078	&0.328	&0.056	&0.867	&15.2	&                                               	&$95$\\
\hline
11 	&Gosgen-38 		 &0.619	&0.067	&0.272	&0.042	&0.955	&5.4	&\rdelim\}{3}{20pt}[2.0] \rdelim\}{4}{20pt}[3.8]	&$37.9$\\
12 	&Gosgen-46 		 &0.584	&0.068	&0.298	&0.050	&0.981	&5.4	&                                               	&$45.9$\\
13 	&Gosgen-65 		 &0.543	&0.070	&0.329	&0.058	&0.915	&6.7	&                                               	&$64.7$\\
14 	&ILL 			 &1	&0	&0	&0	&0.792	&9.1	&                                               	&$8.76$\\
\hline
15 	&Krasnoyarsk87-33 	 &1	&0	&0	&0	&0.925	&5.0	&\rdelim\}{2}{20pt}[4.1]	&$32.8$\\
16 	&Krasnoyarsk87-92 	 &1	&0	&0	&0	&0.942	&20.4	&                       	&$92.3$\\
17 	&Krasnoyarsk94-57 	 &1	&0	&0	&0	&0.936	&4.2	&0                      	&$57$\\
18 	&Krasnoyarsk99-34 	 &1	&0	&0	&0	&0.946	&3.0	&0                      	&$34$\\
\hline
19 	&SRP-18 		 &1	&0	&0	&0	&0.941	&2.8	&0	&$18.2$\\
20 	&SRP-24 		 &1	&0	&0	&0	&1.006	&2.9	&0	&$23.8$\\
\hline
21 	&Nucifer 		 &0.926	&0.061	&0.008	&0.005	&1.014	&10.7	&0	&$7.2$\\
22 	&Chooz 			 &0.496	&0.087	&0.351	&0.066	&0.996	&3.2	&0	&$\approx 1000$\\
23 	&Palo Verde 		 &0.600	&0.070	&0.270	&0.060	&0.997	&5.4	&0	&$\approx 800$\\
24 	&Daya Bay 		 &0.561	&0.076	&0.307	&0.056	&0.946	&2.0	&0	&$\approx 550$\\
25 	&RENO 			 &0.569	&0.073	&0.301	&0.056	&0.946	&2.1	&0	&$\approx 410$\\
26 	&Double Chooz 		 &0.511	&0.087	&0.340	&0.062	&0.935	&1.4	&0	&$\approx 415$\\
\hline
\end{tabular}
\end{center}
\caption{ \label{tab:rat}
List of the experiments which measured the absolute reactor antineutrino flux.
For each experiment numbered with the index $a$,
the index $k = 235, 238, 239, 241$
indicate the four isotopes
$^{235}\text{U}$,
$^{238}\text{U}$,
$^{239}\text{Pu}$, and
$^{241}\text{Pu}$,
$f^{a}_{k}$ are the fission fractions,
$R_{a,\text{SH}}^{\text{exp}}$ is the ratio of measured and predicted rates,
$\sigma_{a}^{\text{exp}}$ is the corresponding relative experimental uncertainty,
$\sigma_{a}^{\text{cor}}$ is the relative systematic uncertainty
which is correlated in each group of experiments indicated by the braces,
and
$L_{a}$ is the source-detector distance.
}
\end{table}

The experimental uncertainties and their correlations listed in Table~\ref{tab:rat}
have been obtained from the corresponding experimental papers.
In particular:
\begin{itemize}

\item
The Bugey-4 and Rovno91 experiments
have a correlated 1.4\% uncertainty,
because they used the same detector
\cite{Declais:1994ma}.

\item
The Rovno88 experiments
have a correlated 2.2\% reactor-related uncertainty
\cite{Afonin:1988gx}.
In addition,
each of the each of the two groups
of integral (Rovno88-1I and Rovno88-2I)
and spectral (Rovno88-1S, Rovno88-2S, and Rovno88-3S)
measurements
have a correlated 3.1\% detector-related uncertainty
\cite{Afonin:1988gx}.

\item
The Bugey-3 experiments
have a correlated 4.0\% uncertainty
obtained from Tab.~9 of \cite{Declais:1994ma}.

\item
The Gosgen and ILL experiments
have a correlated 3.8\% uncertainty,
because they used the same detector
\cite{Zacek:1986cu}.
In addition, the Gosgen experiments
have a correlated 2.0\% reactor-related uncertainty
\cite{Zacek:1986cu}.

\item
The 1987 Krasnoyarsk87-33 and Krasnoyarsk87-92 experiments
have a correlated 4.1\% uncertainty, because they used the same detector
at 32.8 and 92.3 m from two reactors \cite{Vidyakin:1987ue}.
The Krasnoyarsk94-57 experiment was performed in 1990-94 with a different detector at
57.0 and 57.6 m from the same two reactors
\cite{Vidyakin:1990iz}.
The Krasnoyarsk99-34 experiment was performed in 1997-99 with a new integral-type detector
at 34 m from the same reactor of the Krasnoyarsk87-33 experiment
\cite{Kozlov:1999cs}.
There may be reactor-related uncertainties correlated among the four
Krasnoyarsk experiments,
but,
taking into account the time separations and the absence of any information,
we conservatively neglected them.

\item
Following Ref.~\cite{Zhang:2013ela},
we considered the two SRP measurements
as uncorrelated,
because the two measurements would be incompatible
with the correlated uncertainty estimated in Ref.~\cite{Greenwood:1996pb}.

\end{itemize}

In order to investigate which
of the fluxes of the fissile isotopes is responsible for the anomaly,
we consider the theoretical ratios
\begin{equation}
R_{a}^{\text{th}}
=
\dfrac{\sum_{k} f^{a}_{k} r_{k} \sigma_{f,k}^{\text{SH}}}{\sum_{k} f^{a}_{k} \sigma_{f,k}^{\text{SH}}}
,
\label{rth}
\end{equation}
where
the coefficient $r_{k}$
is the needed correction for the
flux of the $k$ fissile isotope.
We derive the values of the coefficients $r_{k}$
by fitting the experimental ratios
$R_{a,\text{SH}}^{\text{exp}}$
with the least-squares function
\begin{equation}
\chi^2
=
\sum_{a,b}
\left( R_{a}^{\text{th}} - R_{a,\text{SH}}^{\text{exp}} \right)
\left( V^{-1} \right)_{ab}
\left( R_{b}^{\text{th}} - R_{b,\text{SH}}^{\text{exp}} \right)
,
\label{chi}
\end{equation}
where $V$ is the covariance matrix constructed with the uncertainties in Table~\ref{tab:rat}.

\begin{figure*}[t!]
\centering
\setlength{\tabcolsep}{0pt}
\begin{tabular}{cc}
\subfigure[]{\label{fig:plt0}
\includegraphics*[width=0.49\linewidth]{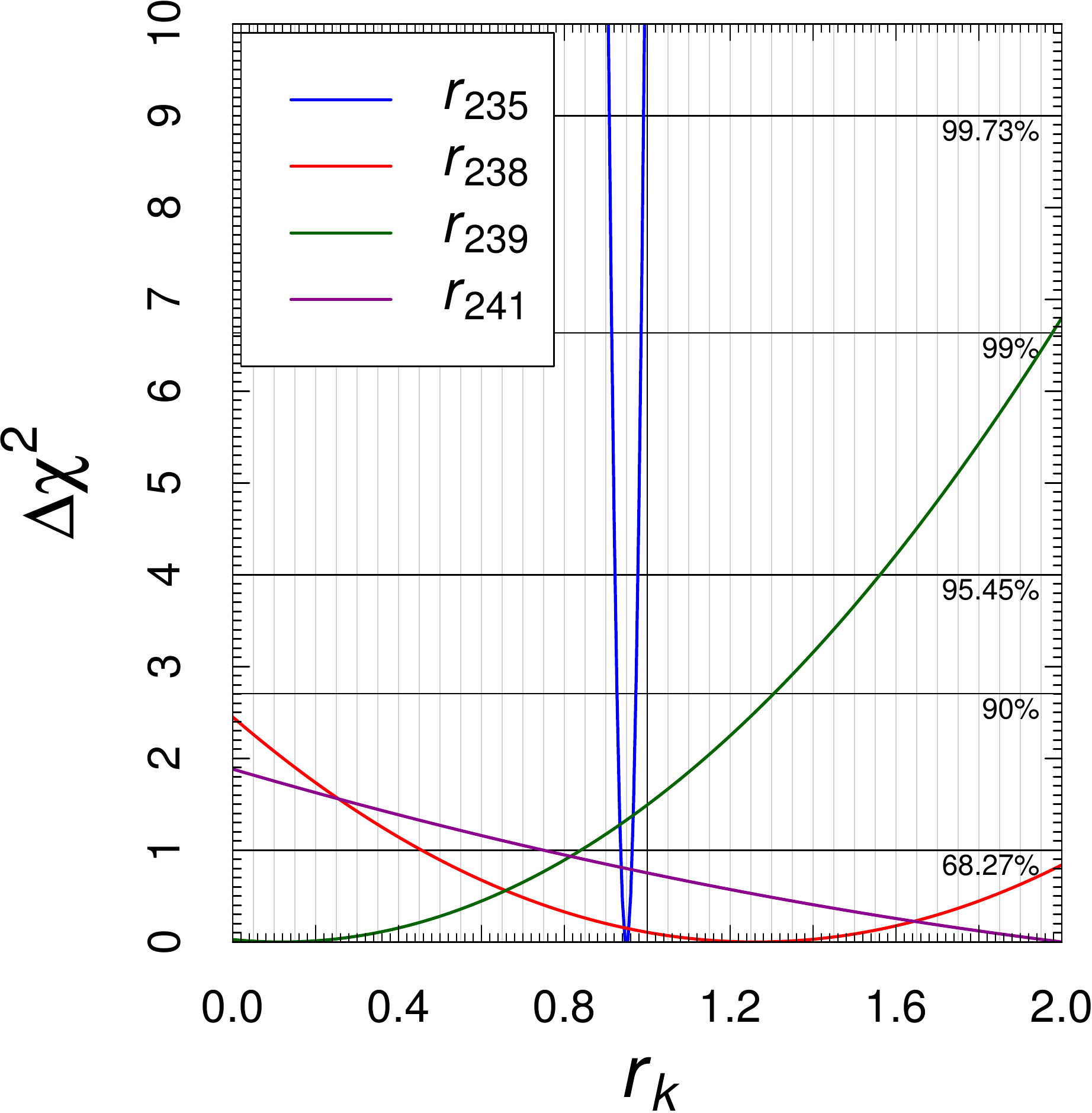}
}
&
\subfigure[]{\label{fig:plt1}
\includegraphics*[width=0.49\linewidth]{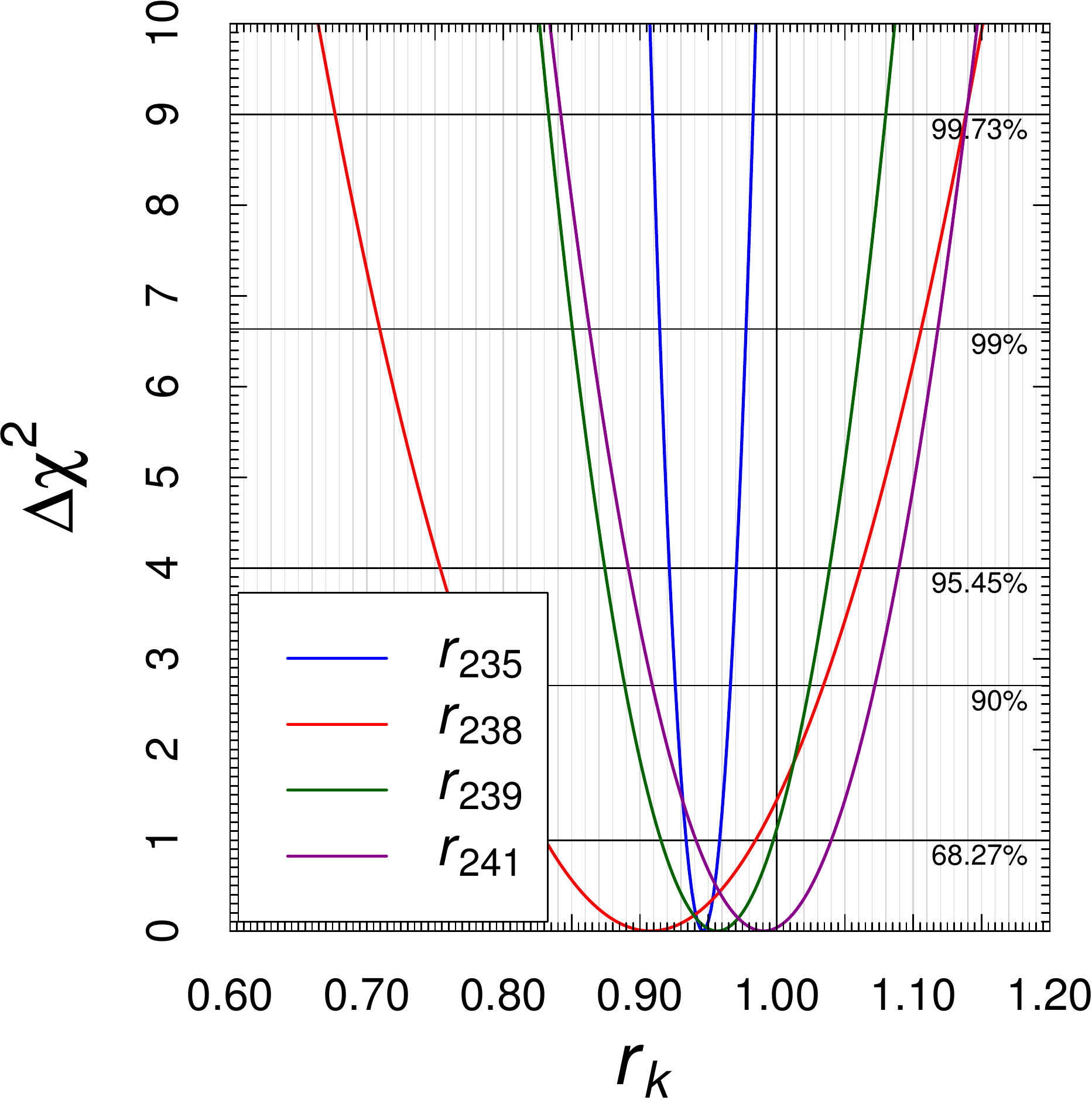}
}
\end{tabular}
\caption{ \label{fig:plt}
Marginal $\Delta\chi^2 = \chi^2 - \chi^2_{\text{min}}$
for the coefficients $r_{k}$ of the four antineutrino fluxes
obtained from the fit of the reactor antineutrino data in Table~\ref{tab:rat}
with the least-squares function in Eq.~(\ref{chi}) \subref{fig:plt0}
and that in Eq.~(\ref{chitilde}) \subref{fig:plt1}.
}
\end{figure*}

The fit of the data in Table~\ref{tab:rat} gives
$\chi^2_{\text{min}} = 16.5$
with
$22$
degrees of freedom,
which correspond to an excellent
$78\%$
goodness of fit.
On the other hand,
the null hypothesis (all $r_{k}=1$) has
$\chi^2 = 97.8$
with
$26$
degrees of freedom,
which corresponds to a disastrous
%$0.00000003\%$
goodness of fit.

Figure~\ref{fig:plt0} shows the marginal
$\Delta\chi^2 = \chi^2 - \chi^2_{\text{min}}$
for the coefficients $r_{k}$ of the four antineutrino fluxes
obtained from the fit.
One can see that the values of
$r_{238}$,
$r_{239}$, and
$r_{241}$
are not sharply constrained:
they are compatible with unity,
but significantly different values are allowed.
On the other hand,
$r_{235}$
is sharply determined by the data:
\begin{equation}
r_{235}
=
0.950
\pm
0.014
\quad
\Rightarrow
\quad
\sigma_{f,235}
=
r_{235}
\sigma_{f,235}^{\text{SH}}
=
(
6.35
\pm
0.09
)
\times
10^{-43} \, \text{cm}^2 / \text{fission}
.
\label{r235}
\end{equation}
This value of the $^{235}\text{U}$ cross section per fission
must be compared with the calculated value in Table~\ref{tab:csf}:
$\displaystyle
\sigma_{f,235}^{\text{SH}}
=
(
6.69
\pm
0.14
)
\times
10^{-43} \, \text{cm}^2 / \text{fission}
$.
The value of $\sigma_{f,235}$
obtained from the fit has an uncertainty that is smaller than
the uncertainty of
$\sigma_{f,235}^{\text{SH}}$.
Adding the two uncertainties quadratically,
there is a discrepancy of
$2.0\sigma$
between the two values.

However,
one can question the reliability of the calculation above
by noting that the large deviations from unity of the best-fit values
$r_{239}^{\text{bf}} = 0.118$
and
$r_{241}^{\text{bf}} = 3.490$,
are excessive for a physical explanation.
In order to restrict the values of
$r_{238}$,
$r_{239}$, and
$r_{241}$
to reasonable intervals around unity,
we use the least-squares function
\begin{equation}
\widetilde{\chi}^2
=
\chi^2
+
\sum_{k}
\left( \dfrac{1-r_{k}}{\Delta{r}_{k}} \right)^2
.
\label{chitilde}
\end{equation}
Taking into account the 5\% uncertainty of the reactor neutrino flux recently advocated in Refs.~\cite{Vogel:2016ted,Hayes:2016qnu,Huber-Nu2016},
we consider
$\Delta{r}_{235}=\Delta{r}_{239}=\Delta{r}_{241}=0.05$,
and
we slightly increase the large uncertainty of $r_{238}$ in Table~\ref{tab:csf}
by considering
$\Delta{r}_{238}=0.1$.
The results of the fit are shown in Fig.~\ref{fig:plt1}.
One can see that the values of all the ratios are
now in a reasonable range around unity:
\begin{equation}
r_{235}
=
0.946
\pm
0.012
,
\quad
r_{238}
=
0.908
\pm
0.077
,
\quad
r_{239}
=
0.956
\pm
0.041
,
\quad
r_{241}
=
0.990
\pm
0.049
.
\label{rn}
\end{equation}
The values of
$r_{238}$,
$r_{239}$, and
$r_{241}$
have still large uncertainties and they are compatible with unity.
For $r_{235}$ we obtain a result similar to that in Eq.~(\ref{r235}),
but more reliable,
because of the more reasonable values of
$r_{238}$,
$r_{239}$, and
$r_{241}$.
In this case,
for $\sigma_{f,235}$
we obtain
\begin{equation}
\sigma_{f,235}
=
(
6.33
\pm
0.08
)
\times
10^{-43} \, \text{cm}^2 / \text{fission}
.
\label{csf235n}
\end{equation}
There is now a discrepancy of
$2.2\sigma$
with the calculated value $\sigma_{f,235}^{\text{SH}}$
in Table~\ref{tab:csf}.

In conclusion,
we obtained the reliable precise determination in Eq.~(\ref{csf235n})
of the $^{235}\text{U}$ cross section per fission
which is more precise than the
calculated value and differs from it
by $2.2\sigma$.
Hence,
if the reactor neutrino anomaly is due to a miscalculation of the antineutrino fluxes,
it is very likely that at least the calculation of the $^{235}\text{U}$ flux
must be revised.

%\bibliographystyle{iopart-num}
%\bibliography{aap2016-pro}

\end{document}